\documentclass[twocolumn]{aastex631}

\usepackage{amsmath}
\usepackage{hyperref}
\usepackage{xcolor}

\newcommand{\allskyname}{time-dependent~GFU}
\newcommand{\mytablenotemark}[1]{\scriptsize \rm \color{blue} \raisebox{0.3em}{#1}}
\newcommand{\mytablenotetext}[1]{\tablenotetext{ \rm \color{blue} #1}}
\newcommand{\sy}[1]{#1}

\begin{document}

\affiliation{III. Physikalisches Institut, RWTH Aachen University, D-52056 Aachen, Germany}
\affiliation{Department of Physics, University of Adelaide, Adelaide, 5005, Australia}
\affiliation{Dept. of Physics and Astronomy, University of Alaska Anchorage, 3211 Providence Dr., Anchorage, AK 99508, USA}
\affiliation{Dept. of Physics, University of Texas at Arlington, 502 Yates St., Science Hall Rm 108, Box 19059, Arlington, TX 76019, USA}
\affiliation{School of Physics and Center for Relativistic Astrophysics, Georgia Institute of Technology, Atlanta, GA 30332, USA}
\affiliation{Dept. of Physics, Southern University, Baton Rouge, LA 70813, USA}
\affiliation{Dept. of Physics, University of California, Berkeley, CA 94720, USA}
\affiliation{Lawrence Berkeley National Laboratory, Berkeley, CA 94720, USA}
\affiliation{Institut f{\"u}r Physik, Humboldt-Universit{\"a}t zu Berlin, D-12489 Berlin, Germany}
\affiliation{Fakult{\"a}t f{\"u}r Physik {\&} Astronomie, Ruhr-Universit{\"a}t Bochum, D-44780 Bochum, Germany}
\affiliation{Universit{\'e} Libre de Bruxelles, Science Faculty CP230, B-1050 Brussels, Belgium}
\affiliation{Vrije Universiteit Brussel (VUB), Dienst ELEM, B-1050 Brussels, Belgium}
\affiliation{Dept. of Physics, Simon Fraser University, Burnaby, BC V5A 1S6, Canada}
\affiliation{Department of Physics and Laboratory for Particle Physics and Cosmology, Harvard University, Cambridge, MA 02138, USA}
\affiliation{Dept. of Physics, Massachusetts Institute of Technology, Cambridge, MA 02139, USA}
\affiliation{Dept. of Physics and The International Center for Hadron Astrophysics, Chiba University, Chiba 263-8522, Japan}
\affiliation{Department of Physics, Loyola University Chicago, Chicago, IL 60660, USA}
\affiliation{Dept. of Physics and Astronomy, University of Canterbury, Private Bag 4800, Christchurch, New Zealand}
\affiliation{Dept. of Physics, University of Maryland, College Park, MD 20742, USA}
\affiliation{Dept. of Astronomy, Ohio State University, Columbus, OH 43210, USA}
\affiliation{Dept. of Physics and Center for Cosmology and Astro-Particle Physics, Ohio State University, Columbus, OH 43210, USA}
\affiliation{Niels Bohr Institute, University of Copenhagen, DK-2100 Copenhagen, Denmark}
\affiliation{Dept. of Physics, TU Dortmund University, D-44221 Dortmund, Germany}
\affiliation{Dept. of Physics and Astronomy, Michigan State University, East Lansing, MI 48824, USA}
\affiliation{Dept. of Physics, University of Alberta, Edmonton, Alberta, T6G 2E1, Canada}
\affiliation{Erlangen Centre for Astroparticle Physics, Friedrich-Alexander-Universit{\"a}t Erlangen-N{\"u}rnberg, D-91058 Erlangen, Germany}
\affiliation{Physik-department, Technische Universit{\"a}t M{\"u}nchen, D-85748 Garching, Germany}
\affiliation{D{\'e}partement de physique nucl{\'e}aire et corpusculaire, Universit{\'e} de Gen{\`e}ve, CH-1211 Gen{\`e}ve, Switzerland}
\affiliation{Dept. of Physics and Astronomy, University of Gent, B-9000 Gent, Belgium}
\affiliation{Dept. of Physics and Astronomy, University of California, Irvine, CA 92697, USA}
\affiliation{Karlsruhe Institute of Technology, Institute for Astroparticle Physics, D-76021 Karlsruhe, Germany}
\affiliation{Karlsruhe Institute of Technology, Institute of Experimental Particle Physics, D-76021 Karlsruhe, Germany}
\affiliation{Dept. of Physics, Engineering Physics, and Astronomy, Queen's University, Kingston, ON K7L 3N6, Canada}
\affiliation{Department of Physics {\&} Astronomy, University of Nevada, Las Vegas, NV 89154, USA}
\affiliation{Nevada Center for Astrophysics, University of Nevada, Las Vegas, NV 89154, USA}
\affiliation{Dept. of Physics and Astronomy, University of Kansas, Lawrence, KS 66045, USA}
\affiliation{Centre for Cosmology, Particle Physics and Phenomenology - CP3, Universit{\'e} catholique de Louvain, Louvain-la-Neuve, Belgium}
\affiliation{Department of Physics, Mercer University, Macon, GA 31207-0001, USA}
\affiliation{Dept. of Astronomy, University of Wisconsin{\textemdash}Madison, Madison, WI 53706, USA}
\affiliation{Dept. of Physics and Wisconsin IceCube Particle Astrophysics Center, University of Wisconsin{\textemdash}Madison, Madison, WI 53706, USA}
\affiliation{Institute of Physics, University of Mainz, Staudinger Weg 7, D-55099 Mainz, Germany}
\affiliation{Department of Physics, Marquette University, Milwaukee, WI 53201, USA}
\affiliation{Institut f{\"u}r Kernphysik, Universit{\"a}t M{\"u}nster, D-48149 M{\"u}nster, Germany}
\affiliation{Bartol Research Institute and Dept. of Physics and Astronomy, University of Delaware, Newark, DE 19716, USA}
\affiliation{Dept. of Physics, Yale University, New Haven, CT 06520, USA}
\affiliation{Columbia Astrophysics and Nevis Laboratories, Columbia University, New York, NY 10027, USA}
\affiliation{Dept. of Physics, University of Oxford, Parks Road, Oxford OX1 3PU, United Kingdom}
\affiliation{Dipartimento di Fisica e Astronomia Galileo Galilei, Universit{\`a} Degli Studi di Padova, I-35122 Padova PD, Italy}
\affiliation{Dept. of Physics, Drexel University, 3141 Chestnut Street, Philadelphia, PA 19104, USA}
\affiliation{Physics Department, South Dakota School of Mines and Technology, Rapid City, SD 57701, USA}
\affiliation{Dept. of Physics, University of Wisconsin, River Falls, WI 54022, USA}
\affiliation{Dept. of Physics and Astronomy, University of Rochester, Rochester, NY 14627, USA}
\affiliation{Department of Physics and Astronomy, University of Utah, Salt Lake City, UT 84112, USA}
\affiliation{Dept. of Physics, Chung-Ang University, Seoul 06974, Republic of Korea}
\affiliation{Oskar Klein Centre and Dept. of Physics, Stockholm University, SE-10691 Stockholm, Sweden}
\affiliation{Dept. of Physics and Astronomy, Stony Brook University, Stony Brook, NY 11794-3800, USA}
\affiliation{Dept. of Physics, Sungkyunkwan University, Suwon 16419, Republic of Korea}
\affiliation{Institute of Basic Science, Sungkyunkwan University, Suwon 16419, Republic of Korea}
\affiliation{Institute of Physics, Academia Sinica, Taipei, 11529, Taiwan}
\affiliation{Dept. of Physics and Astronomy, University of Alabama, Tuscaloosa, AL 35487, USA}
\affiliation{Dept. of Astronomy and Astrophysics, Pennsylvania State University, University Park, PA 16802, USA}
\affiliation{Dept. of Physics, Pennsylvania State University, University Park, PA 16802, USA}
\affiliation{Dept. of Physics and Astronomy, Uppsala University, Box 516, SE-75120 Uppsala, Sweden}
\affiliation{Dept. of Physics, University of Wuppertal, D-42119 Wuppertal, Germany}
\affiliation{Deutsches Elektronen-Synchrotron DESY, Platanenallee 6, D-15738 Zeuthen, Germany}

\author[0000-0001-6141-4205]{R. Abbasi}
\affiliation{Department of Physics, Loyola University Chicago, Chicago, IL 60660, USA}

\author[0000-0001-8952-588X]{M. Ackermann}
\affiliation{Deutsches Elektronen-Synchrotron DESY, Platanenallee 6, D-15738 Zeuthen, Germany}

\author{J. Adams}
\affiliation{Dept. of Physics and Astronomy, University of Canterbury, Private Bag 4800, Christchurch, New Zealand}

\author[0000-0002-9714-8866]{S. K. Agarwalla}
\altaffiliation{also at Institute of Physics, Sachivalaya Marg, Sainik School Post, Bhubaneswar 751005, India}
\affiliation{Dept. of Physics and Wisconsin IceCube Particle Astrophysics Center, University of Wisconsin{\textemdash}Madison, Madison, WI 53706, USA}

\author[0000-0003-2252-9514]{J. A. Aguilar}
\affiliation{Universit{\'e} Libre de Bruxelles, Science Faculty CP230, B-1050 Brussels, Belgium}

\author[0000-0003-0709-5631]{M. Ahlers}
\affiliation{Niels Bohr Institute, University of Copenhagen, DK-2100 Copenhagen, Denmark}

\author[0000-0002-9534-9189]{J.M. Alameddine}
\affiliation{Dept. of Physics, TU Dortmund University, D-44221 Dortmund, Germany}

\author{N. M. Amin}
\affiliation{Bartol Research Institute and Dept. of Physics and Astronomy, University of Delaware, Newark, DE 19716, USA}

\author[0000-0001-9394-0007]{K. Andeen}
\affiliation{Department of Physics, Marquette University, Milwaukee, WI 53201, USA}

\author[0000-0003-4186-4182]{C. Arg{\"u}elles}
\affiliation{Department of Physics and Laboratory for Particle Physics and Cosmology, Harvard University, Cambridge, MA 02138, USA}

% ...

\date{\today}

\collaboration{427}{IceCube Collaboration}

\title{Search for neutrino doublets and triplets using 11.4 years of IceCube data}

\begin{abstract}
\sy{We report a search for high-energy astrophysical neutrino multiplets,
detections of multiple neutrino clusters in the same direction
within 30 days, based on an analysis of 11.4 years of IceCube data. 
A new search method optimized for transient neutrino emission with a monthly time scale
is employed, providing a higher sensitivity to neutrino fluxes.
This result is sensitive to neutrino transient emission, reaching per-flavor flux 
of approximately $10^{-10}\ {\rm erg}\ {\rm cm}^{-2}\ {\rm sec}^{-1}$
from the Northern sky in the energy range $E\gtrsim 50$~TeV.
The number of doublets and triplets identified in this search is compatible with the atmospheric 
background hypothesis, which leads us to set limits on the nature of 
neutrino transient sources with 
emission timescales of one month. 
}

\end{abstract}

\keywords{Neutrino astronomy(1100) --- Neutrino telescopes(1105) --- Optical astronomy(1776)}


\section{Introduction}\label{sec:intro}

The IceCube Neutrino Observatory discovered a diffuse flux of 
high-energy astrophysical neutrinos in 2013~\citep{PhysRevLett.111.021103}. The observed 
flux per neutrino flavor is at the level of $\phi_{\nu}\sim 10^{-18}~\mathrm{GeV}^{-1}\,\mathrm{cm}^{-2}\,\mathrm{sec}^{-1}\,\mathrm{sr}^{-1}$ 
at $E_\nu=\mathrm{100}~\mathrm{TeV}$ with a spectral 
index between 2.4 and 2.9 depending on various neutrino selections~\citep{Abbasi_2022, PhysRevLett.125.121104_cascade_nuenutau, Aartsen:456632_glashow, PhysRevD.104.022002_HESE}. Hereafter, a flux implicitly represents the sum of neutrinos and antineutrinos unless otherwise stated.

The origin of the diffuse flux is uncertain, but there are various hints from 
the studies of 
the individual point source searches. 
In 2017, the association of a real-time neutrino alert (IceCube-170922A) 
with a high-energy gamma-ray flare observed by Fermi-LAT~\citep{doi:10.1126/science.aat1378} suggested that 
a blazar TXS~0506+056 was likely a source of high-energy neutrinos. 
However, blazars cannot explain all of the diffuse flux observed by IceCube~\citep{Aartsen_2017_BLLAC}.
In 2022, the nearby active galaxy NGC~1068 was 
identified as a neutrino source by an excess 
of neutrinos with energies of $1.5\,\mathchar`-\,15~\mathrm{TeV}$ above 
backgrounds with significance of 4.2$\sigma$~\citep{scienceNGC1068}. 
In 2023, neutrino emission from the galactic plane was 
reported at 4.5$\sigma$ significance~\citep{scienceGP2023}, 
but this only contributes to $6\,\mathchar`-\,13$\% of the isotropic diffuse flux at 30~TeV.

\sy{Astrophysical transient phenomena with a time scale of weeks to months have been proposed
as neutrino sources.}
For example, a subclass of \sy{core-collapse supernovae} (CCSNe) radiates $\mathcal{O}(100)$ brighter than 
standard SNe, possibly due to interactions between the ejecta 
and dense circumstellar materials~\citep{Moriya_SLSN, annurev_SLSN}.
In such environments, along with the prompt emission of MeV energy neutrinos, 
efficient generation of 1-100~TeV neutrinos can occur over a timescale of 
up to 100 days~\citep{PhysRevD.84.043003_CCSN_CSM, Kheirandish_2023, 10.1093/mnras/stad2025, PhysRevD.97.081301_SNCSM}. 
Tidal disruption events (TDEs) are observed 
by their lasting several months emission of optical radiation. 
High-energy neutrino emission from TDEs was 
predicted to react a total bolometric energy of $\mathcal{O}(10^{54})~\mathrm{erg}~$\citep{Hayasaki_2019, Murase_2020_TDE, Winter_2023_TDE}.
Furthermore, \cite{c82b900f577742bab62febd116bdf4fe} reported a 
possible association between IceCube real-time neutrino alerts and TDE flares.

A search for multiple neutrinos originating from the same direction within a timescale of one month 
is a straightforward method for capturing these phenomena. 
If these transient sources contribute to the diffuse neutrino flux, 
the detection of neutrino {\it multiplets} 
can identify sources and extract the characteristics of the source population.
For transient sources of burst rate density $\rho_\nu$ with all-flavor neutrino isotropic 
emission energy per source, $\mathcal{E}_\nu$, the diffuse flux scales approximately 
as $\displaystyle E_\nu^2\phi_\nu \propto \mathcal{E}_\nu\rho_\nu$ while the detection 
rate of doublets is approximately proportional to $\mathcal{E}_\nu^2\rho_\nu$. Thus, 
the search for doublets (and multiplets in general) is more sensitive to
a population of bright ({\it i.e.,} higher $\mathcal{E}_\nu$), and rarer ({\it i.e.,} lower $\rho_\nu$)
transient sources. Furthermore, a null detection of astrophysical
neutrino multiplets constrains these parameters and 
can rule out candidate transient sources for the diffuse flux~\citep{KOWALSKI2007533, PhysRevLett.122.051102_shortmultiplet, Yoshida_2022}.

\sy{
We report here a search for neutrino doublets and triplets using IceCube data.
The search scheme is optimized for a relatively long transient time scale of 30~days. 
This is 
different from our previous doublet search~\citep{PhysRevLett.122.051102_shortmultiplet}, 
where the search was conducted for short transient with less than 100~s, 
corresponding to a much faster emission process 
such as gamma-ray bursts (GRBs).
}

The remainder of this paper is organized as follows. 
Sec.~\ref{icecube_evsel} explains IceCube observatory 
and the selection method 
for neutrino events for this analysis. 
In Sec.~\ref{Multipletana} presents the analysis method of the 
multiplets, together with its sensitivity.
Sec.~\ref{result_sec} presents the results of our analysis of the dataset. 
Sec.~\ref{discussion} discusses the outcomes of the main results.
We also report a correlation analysis between multiplets 
and X-ray data recorded by the Monitor of All-sky X-ray Images (MAXI). 
Finally, in Sec.~\ref{conclusion_sec}, we summarize the conclusions of the search for multiplets.

\section{The IceCube observatory\label{icecube_evsel}}
The IceCube observatory at the South Pole is a 
cubic-kilometer detector for astrophysical neutrinos~\citep{Aartsen_2017_detector, ABBASI2009294_DAQ}. 
Cherenkov emission from charged particles generated by 
the interaction of high-energy neutrinos are observed using 5160 digital optical modules (DOMs). 
The DOMs host 10" photomultiplier tubes in a high-pressure-resistant 
vessel~\citep{ABBASI2010139_PMT} and were deployed between 1450 and 2450~m below the surface of the 
ice on 86 vertical strings. The detected photon signals are sent to 
the surface IceCube laboratory, and recorded when the trigger condition is satisfied.
The patterns of photons detected from neutrino events are predominantly classified as either {\it tracks} or {\it cascades}. 
A track event can be initiated by the charged current interaction of $\nu_\mu$, and the photons disperse along the path 
of the outgoing muon, providing good angular resolution, 
typically less than $1^\circ$ for high energy ($\gtrsim 10~\mathrm{TeV}$) neutrinos. 
A cascade event is initiated by charged-current interactions of $\nu_e$ and $\nu_\tau$, or 
neutral-current interactions of all neutrino flavors. 
Its photon distribution exhibits more isotropic quasi-pointlike patterns, resulting in a 
poorer resolution than that of a track event.

The event set used in this study is referred to as the {\it GFU sample}~\citep{Kintscher2020Rapid,gfu_2016}, 
\sy{which contains high-quality track events with excellent pointing accuracy}.
The dataset was collected from the completion of IceCube in 2011 until the end of 2022, 
corresponding to a total live time of 11.4 years when accounting for data quality selections. 
The total number of selected events is $N=1,212,746$, and the 
event rate is $3.4~\mathrm{mHz}$.
The sample is also used to issue single high-energy 
track neutrino alerts~\citep{Abbasi_2023}. Neutrino events are further selected 
based on their reconstructed muon energies~\citep{ABBASI2013190_truncated} and directions~\citep{AHRENS2004169_splineMPE}.
This study utilized only the track type 
events in the northern direction with $\delta>-5^{\circ}$, forming the {\it Northern Sky GFU} sample. 
The main background originates from atmospheric neutrinos and dominates by 99.7\%. 
At the South Pole, a one-to-one relationship exists between the declination $\delta$ and the 
zenith angle which allows us to characterize the atmospheric background as a function of $\delta$.

\section{Multiplet analysis \label{Multipletana}}

\sy{
In the following, the emission energy 
of individual neutrino sources is bolometrically defined 
in the energy range from 10~TeV to 1~PeV, which is 
represented as $\mathcal{E}^*_\nu$. 
The expected number of neutrinos passing the event selection 
from a source at redshift $z$, $n_\nu(\mathcal{E}^*_\nu, z)$, yields
a probability of multiplet detection as 
${\epsilon(\mathcal{E}^*_\nu, z) = 1-e^{-n_\nu}-n_\nu e^{-n_\nu}}$.
The total number of 
multiplets (doublets, triplets, quadruplets, $\ldots$, and so on) 
expected 
from the population 
of such sources across the Universe is given by~\citep{Murase:2016gly,Yoshida_2022}:
\begin{align}
N_{\rm M} = \rho_\nu T_\mathrm{obs}  \int dV \epsilon(\mathcal{E}^*_\nu, z) \psi(z),
\end{align}
where $T_\mathrm{obs}$ is a total live time of observation 
and $dV$ is a volume element. 
The factor $\psi(z)$ represents an evolution of the source 
population with redshifts, and in the present analysis,
$\psi(z)$ follows $(1+z)^{3.4}$ for $z\leq 1$ and is constant beyond $z=1$.
This is comparable to the star formation rate (SFR).
For instance, a source population with 
$\mathcal{E}_\nu^* = 10^{52}$~erg and $\rho_\nu=10^{-8}$~Mpc$^{-3}$ yr$^{-1}$
yields $N_{\rm M}\sim 3$ per year for the Northern-sky GFU sample. 
This number is much lower than the number of combinations of 
neutrinos due to atmospheric backgrounds (the rate is $\sim \mathcal{O}(10^5)$ per year).
It is necessary to optimize the search method for such a low detection rate of multiplets. 
Below, we describe our newly developed multiplet finding algorithms and the resultant likelihood construction.
}

\subsection{Selection of event clusters \label{clustering_sec}}

\sy{
Any excess of neutrino events above the atmospheric background from a given patch of
the sky can be identified by evaluating likelihoods of, both, astrophysical and background hypotheses 
for the observed 
events in a sliding search time window.
A previous time-dependent point-source search~\citep{Kintscher2020Rapid} 
(hereafter referred to as {\it \allskyname}) is based on this method 
using the nested-likelihood technique~\citep{BRAUN2008299}.
}

To reduce computation time, we employ a {\it seeded clustering} method in this analysis.
The {\it i}-th event in the sample seeds a search for clustered events in time 
up to $T_{\rm max}=30~\mathrm{days}$ prior to the seed event 
and direction within a $3^\circ$ opening angle.
We then determine every combination of two ({\it i.e.}, doublets) and 
three ({\it i.e., } triplet) event pairs. This pre-selection 
method was previously adopted 
for time-independent searches~\citep{AARTSEN201539_autocorrelation}.
We then select the most statistically significant multiplet among all
combinations based on a test statistic introduced below.
Iterations of this clustering procedure are performed for all seeding events 
in the sample, separately for doublets and triplets, that generate 
the potentially interesting series of two pools of doublets and triplets. 
Note that this algorithm 
in general permits more than three neutrino events to be 
multiple pairs of doublets.

The pools of the multiplets are subject to the hypothesis testing by using 
the following likelihood. 
For a given doublet and a triplet in the 
multiplet pools, we define an extended Poisson likelihood of the signal (background) hypothesis as:
\begin{equation}
    \mathcal{L}^{\rm sig (bg)} = 
    \epsilon_\mathrm{sig (bg)} \prod_i^{N=2 ({\rm or 3)}} P^{\rm E}_\mathrm{sig (bg)}(E^{i})P^{\vec{n}}_\mathrm{sig (bg)}(\vec{n}_i),
    \label{eq:likelihood}
\end{equation}
where $E^i$ is the proxy for the neutrino energy~\citep{ABBASI2013190_truncated}, 
$\vec{n}_i$ is the reconstructed direction of an event {\it i}, and 
$P^{\rm E}_\mathrm{sig (bg)}$ and $P^{\vec{n}}_\mathrm{sig (bg)}$ are the energy and directional
probability density functions (PDFs) for the signal (background) hypothesis.
The first term $\epsilon_\mathrm{sig (bg)}$ represents a probability to observe 
multiplets (more than two background events) in $1~\mathrm{deg}^2$
within the time window $T_\mathrm{max} $ at 
a declination in $\delta$, and given by~\citep{Yoshida_2022}
\begin{eqnarray}
 \epsilon_\mathrm{sig}&=&\left(1-e^{-\Delta N_\mathrm{M}}\right)e^{-\Delta \mu_\mathrm{bg}}, \\
 \epsilon_\mathrm{bg}&=&\left(1-e^{-\Delta \mu_\mathrm{bg}}-\Delta \mu_\mathrm{bg}e^{-\Delta \mu_\mathrm{bg}}\right)e^{-\Delta N_\mathrm{M}},
\end{eqnarray}
where $\Delta N_\mathrm{M}=\Delta N_\mathrm{M}(\sin \delta)$ denotes the 
expected number of multiplets, and
$\Delta \mu_\mathrm{bg}=\Delta \mu_\mathrm{bg}(\sin \delta)$ denotes the expected number of 
background events in the same region, solid angle (1~$\mathrm{deg}^2)$, and time window.

\sy{
The expected number of multiplets $N_\mathrm{M}$ depends on the representative 
neutrino emission energy, $\mathcal{E}^*_\nu$, 
and local burst rate density, $\rho_\nu$. We used a standard-candle 
source model in which all sources yielding
multiplet events are identical across the universe, with universal values 
for $\mathcal{E}^*_\nu$ and $\rho_\nu$.
We use $\mathcal{E}^*_\nu=2.1\times 10^{54}$~erg and $\rho_\nu=1.7\times 10^{-10}$~Mpc$^{-3}$ yr$^{-1}$
as the baseline configuration for the likelihood calculations. 
We confirmed that the choice of baseline values does not significantly change the resulting 
sensitivity.
}

\sy{
The energy PDF $P^{\rm E}_\mathrm{sig}$ depends on the neutrino spectral shape.
We assume that the emission spectrum follows a single power law, $E^{-\gamma}$.
The index $\gamma$ is assumed to be fixed at a baseline 
value of $\gamma=2.3$ (a lower end of the measured spectrum of the diffuse flux by IceCube~\citep{Abbasi_2022}). 
The directional PDF $P^{\vec{n}}_\mathrm{sig}$ relies on the
knowledge of the event localization uncertainty. We describe $P^{\vec{n}}_\mathrm{sig}$
by a two-dimensional Gaussian distribution with an error matrix estimated 
by the fit procedure of the directional reconstruction~\citep{AHRENS2004169_splineMPE}.
The background directional PDF $P^{\rm E}_\mathrm{bg}$ is assumed to be flat (constant) over a $3^\circ $ region.
}

\sy{
Each identified doublet or triplet is then assigned to the likelihood ratio test statistic
\begin{equation}
    \Lambda = 2\log\frac{\mathcal{L}^{\rm sig}(\widehat{\vec{n}})}{\mathcal{L}^{\rm bg}},
    \label{eq:TS}
\end{equation}
where $\widehat{\vec{n}}$ is the direction that maximizes the likelihood.
The distribution of $\Lambda$ under the background-only hypothesis yields the p-value 
or the false alarm rate (FAR) for a given $\Lambda$ value. 
The background distribution of $\Lambda$ for doublets and triplets
are obtained by right-ascension scrambling of the control dataset, 
and exhibit excellent agreement with that of doublets and triplets 
in the full dataset as described in Sec.~\ref{result_sec}.
To unite doublets and triplets, we introduce a TS-score 
defined by the $\Lambda$ value corresponding to the smallest 
FAR out of the pool of doublets and triplets for every seed event. 
Thus, the TS-score distribution 
under the background-only hypothesis provides a unified global p-value or 
FAR for every seed event. 
}

\vspace{5mm}
\subsection{Signal efficiency \label{sigeff_sec}\label{performance_sec}}
\begin{figure}[t]
  \centering
  \includegraphics[width=\linewidth,bb=0 0 720 504]{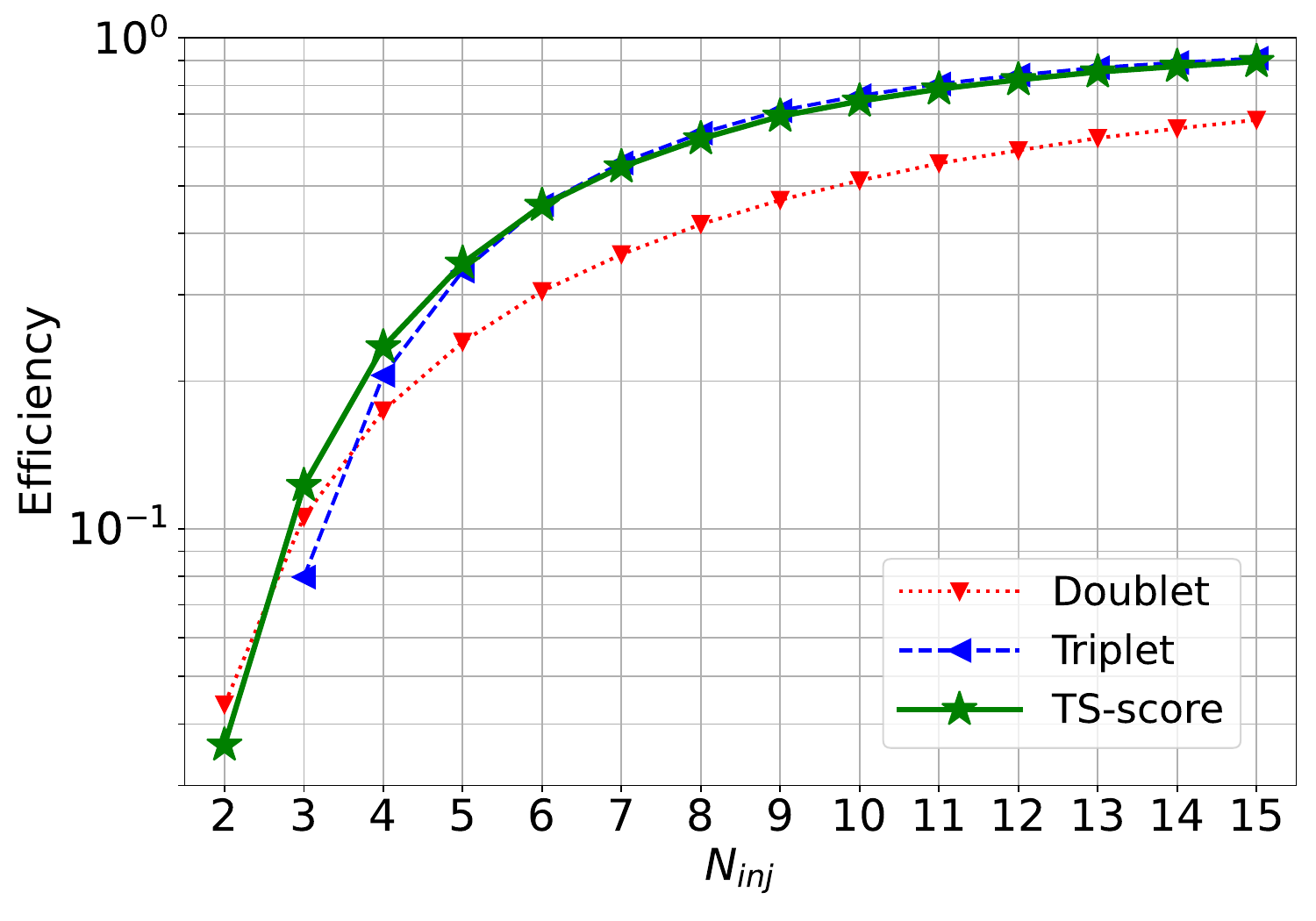}
  \caption{Signal efficiency as a function of injected number of signal neutrino events 
  for doublet (red), triplet (blue), and TS-score (green).
  The definition of the efficiency is given in the main text.}\label{sigeff}
\end{figure}

We derive the signal efficiency as the expected fraction of neutrino transients 
surpassing a TS-score threshold corresponding to $\mathrm{FAR}<1~\mathrm{yr}^{-1}$, 
which represents a metric of the effective performance for a discovery 
for astrophysical multiplet candidates. 
To evaluate the signal efficiency, we simulate transient neutrino emissions 
from hypothetical sources. The number of signal neutrino events 
satisfying the North~Sky GFU selection, $N_\mathrm{inj}$, is introduced here 
as a proxy for the source luminosity. 
The neutrino emission follows an $E^{-2.3}$ spectrum with a flat transient
time profile up to $T_{\rm max}=30$~days from 10~TeV to 1~PeV and zero elsewhere. 
For each signal trial, we {\it inject} 
$N_\mathrm{inj}$ events to the background model prepared by 
right-ascension scrambling of data. 
The injected $N_\mathrm{inj}$ neutrino events seed the 
calculations of $\Lambda$, 
and if any two (three) injected signal events are 
properly selected as a doublet (triplet) with 
corresponding FAR satisfying $< 1~\mathrm{yr}^{-1}$, the trial is defined to surpass the threshold, 
regardless of the consistency between the fit and true injected directions.

Figure~\ref{sigeff} shows the efficiency obtained from these pseudo-experimental trials.
We show the resulting efficiencies for $\mathrm{FAR}<1~\mathrm{yr}^{-1}$ 
by doublet-only, triplet-only, and the combined treatments by 
the TS-score, respectively, for comparison. 
Though the magnitude of efficiency for $N_\mathrm{inj}\leq 5$ is small ($\lesssim $10\%), 
this provides non-negligible opportunities to 
discover neutrino multiplets in the populations of relatively dimmer sources.
For example, for emissions from sources with $\mathcal{E}^*_\nu=10^{52}~\mathrm{erg}$ 
distributed across space, approximately 60 \% of the detection cases range 
from $N_\mathrm{inj}=2$ to 5. 

\subsection{\sy{Sensitivity to a multiplet source flux}}

\begin{figure}[t]
  \centering
  \hspace{-3mm}\includegraphics[width=0.95\linewidth,bb=0 0 648 432]{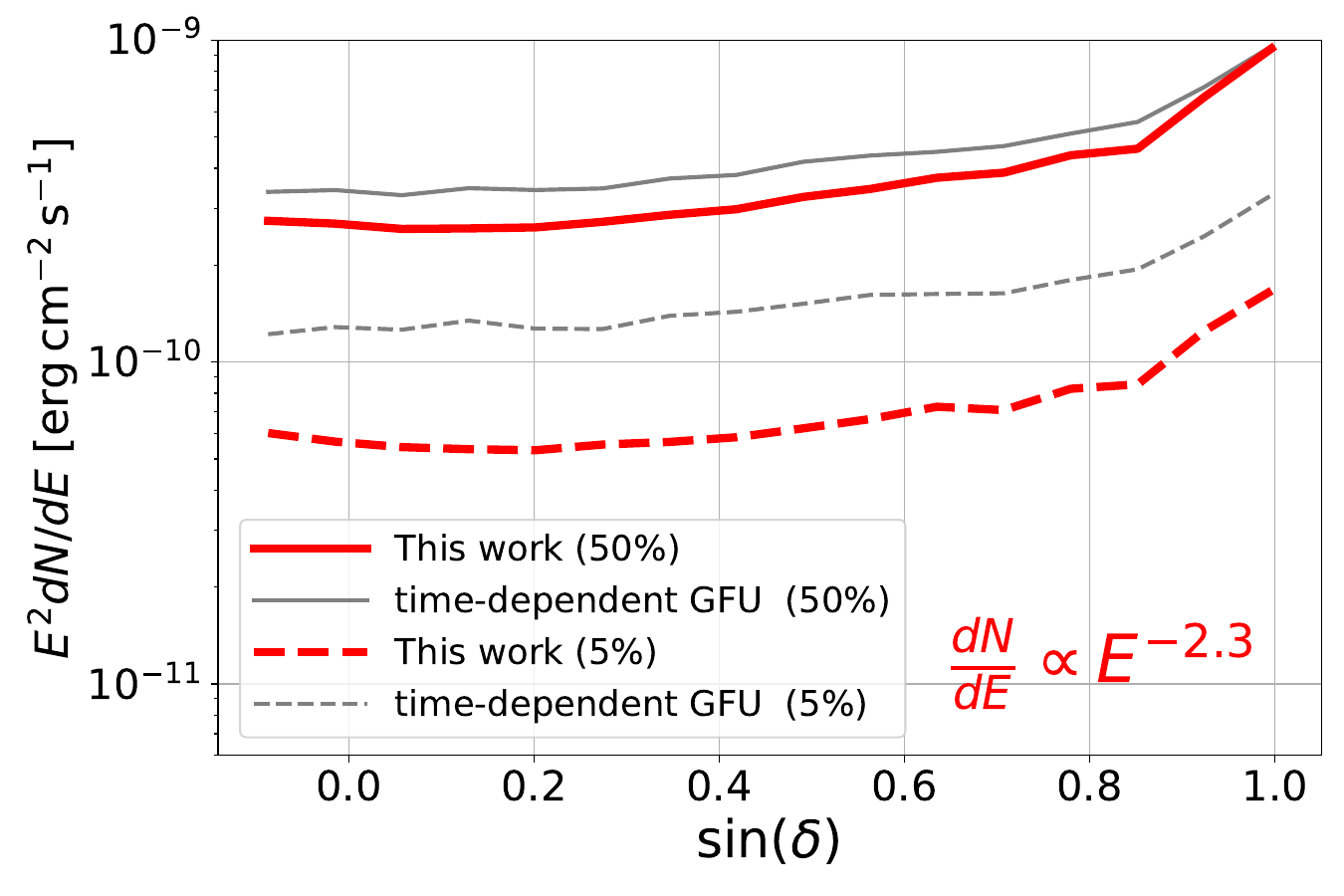}\\
  \caption{Averaged energy flux (sum of $\nu_\mu$ and $\bar{\nu}_\mu$ per flavor) at 100~TeV 
  in $T_\mathrm{max}=30~\mathrm{days}$ observed at a significance corresponding to $\mathrm{FAR}<1~\mathrm{yr}^{-1}$ as a function of $\sin\delta$ for 
  the presented algorithm (red) and \allskyname~(gray) 
  when $\gamma=2.3$. The solid and dashed curves represent probabilities of signal detections of 50\% and 5\%, respectively.}\label{discovery}
\end{figure}

Figure~\ref{discovery} illustrates the sensitivity of the multiplet source flux 
to satisfy $\mathrm{FAR}<1~\mathrm{yr}^{-1}$ as a function of the declination. 
Two cases of 50\% and 5\% efficiencies are 
displayed. When 50\% of efficiency is required (solid), the presented algorithm shows 
compatible performance with the \allskyname ~method, while a factor of $\sim 2$ improvement 
is expected for 5\% efficiency (dashed). 
Because less bright sources should be 
more abundant, the smaller efficiency per source 
could result in a notable number of detections particularly when we 
utilize information of multi-messenger coincidences such as archival 
follow-up analyses or additional observations by alerts.
The improvement comes from the contribution of the small 
number of neutrino detections ($N_\mathrm{inj}\leq 5$) explained in Sec.~\ref{sigeff_sec}.


\subsection{\sy{Main energy range}}

\begin{figure}[h]
  \centering
  \includegraphics[width=0.92\linewidth,bb=0 0 648 504]{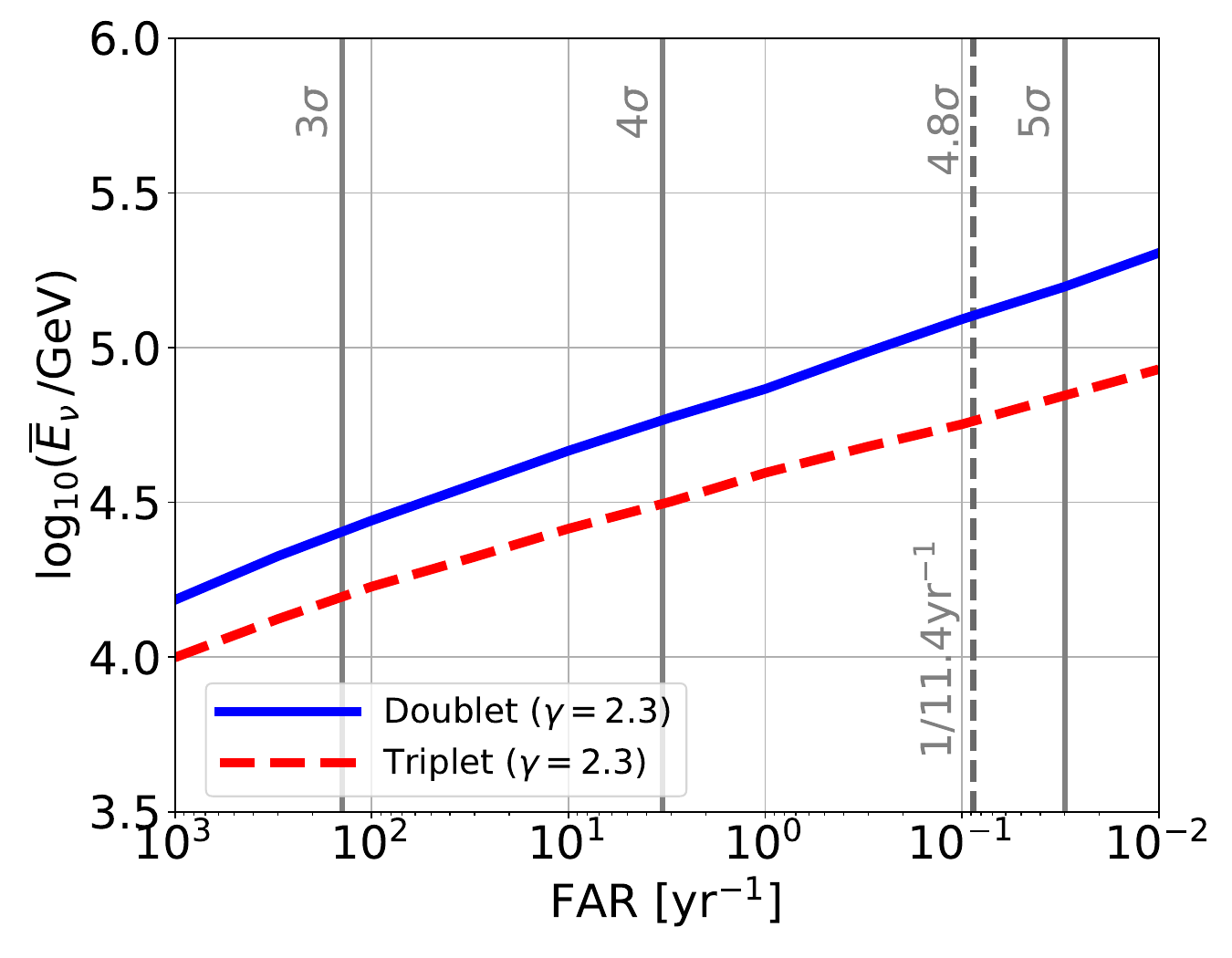}\\
  \caption{Averaged energies of the selected multiplets for various 
  levels of thresholds converted into the corresponding FAR. 
  The energy of the multiplet is the geometric mean of 
  the neutrino energies involved in the multiplets. 
  The solid and dashed curves correspond to doublets and triplets, respectively. 
  The vertical solid lines indicate $3\sigma$, $4\sigma$, and $5\sigma$ of the normalized background 
  distributions, and the dashed-line indicates 4.8$\sigma$ corresponding to 
  $\mathrm{FAR}=1/11.4~\mathrm{yr}$. 
  }\label{energy_selected}
\end{figure}

Figure~\ref{energy_selected} shows the average 
energies of multiplets 
for various thresholds of FAR values. 
At the significance of $4.8\sigma$ of background trials 
(corresponding to $\mathrm{FAR}=1/11.4\mathrm{yr}$), the main neutrino range 
is $E\gtrsim 100$~TeV for doublets and $E\gtrsim 50$~TeV for triplets.

\subsection{Source localization accuracy \label{sec_angres}}

Because of the inclusion of multiple events, the source localization accuracy of multiplets 
is superior to the angular uncertainty of individual events. 
Multiplets whose FARs are less than $1~\mathrm{yr}^{-1}$ results in 
an excellent source localization accuracy of $0.3^\circ$ at 90\% containment. 


\section{Results\label{result_sec}}

\begin{figure}[t]
  \centering
  \hspace{-0mm}\includegraphics[width=1.0\linewidth,bb=0 0 720 504]{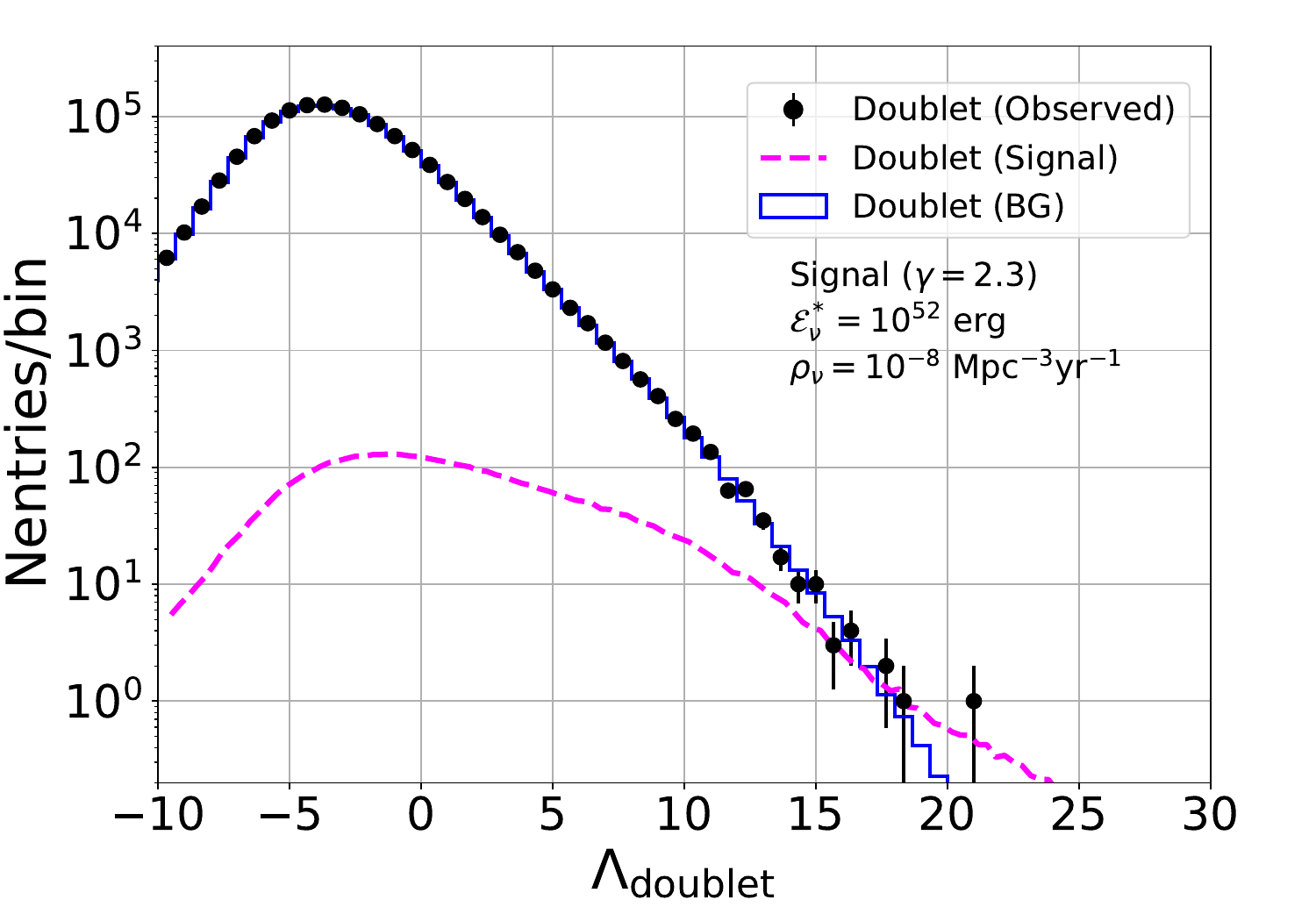}\\
  \vspace{3mm}
  \hspace{-0mm}\includegraphics[width=1.0\linewidth,bb=0 0 720 504]{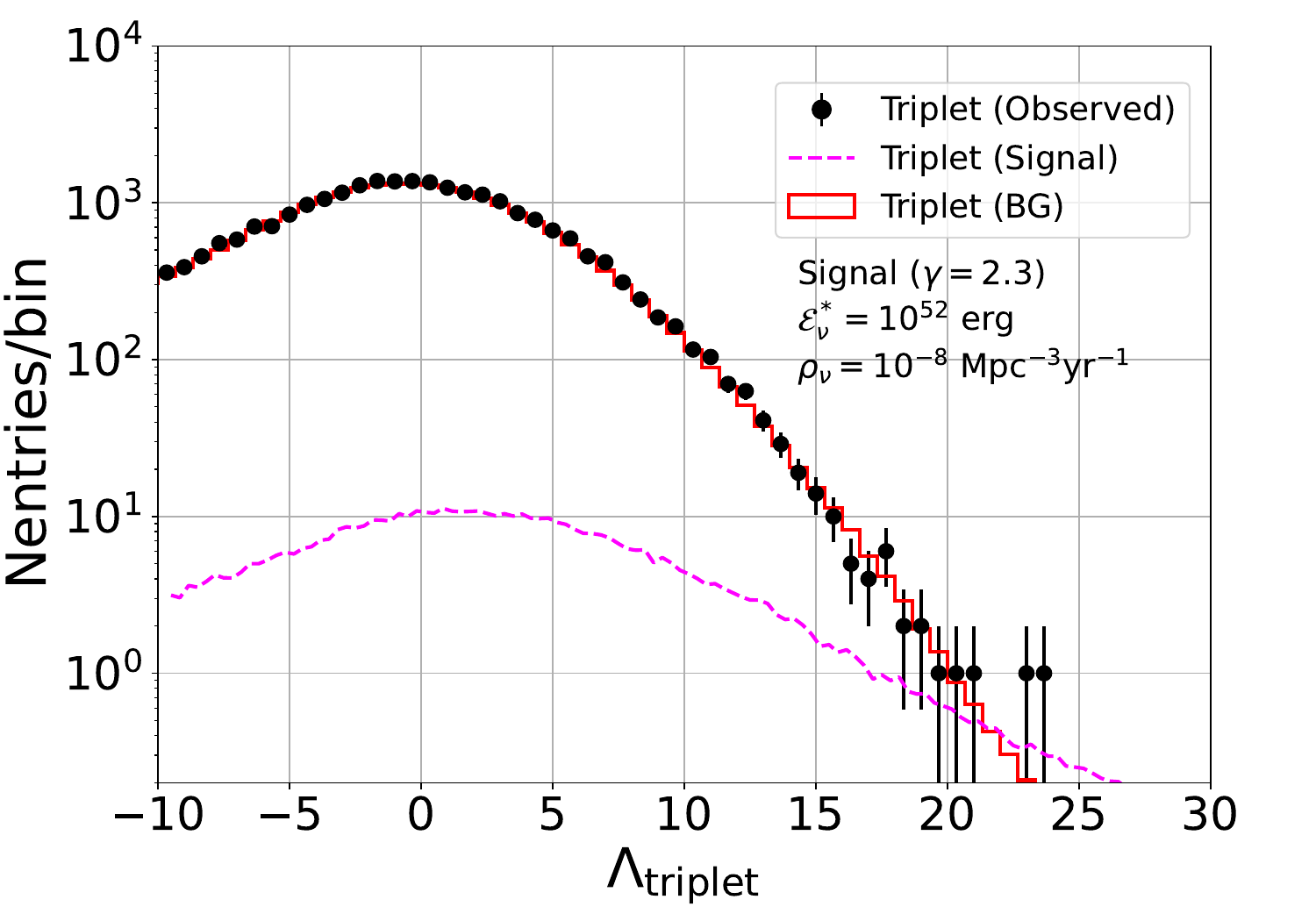}\\
  \caption{Distribution of the test statistics: $\Lambda_\mathrm{doublet}$ (top) and $\Lambda_\mathrm{triplet}$ (bottom). 
 The markers represent observed experimental data, while the solid lines indicate 
 the expected distributions from backgrounds. The magenta 
 dashed lines are the expected distributions of the signal 
 when $\gamma=2.3$, $\mathcal{E}^*_\nu=10^{52}~\mathrm{erg}$, and $\rho_{\nu}=10^{-8}~\mathrm{Mpc}^{-3}\,\mathrm{yr}^{-1}$. These parameters were 
 arbitrary chosen for illustration. }\label{fig_tsdist}
\end{figure}

\begin{figure}[th]
  \hspace{-6mm}\includegraphics[width=1.14\linewidth,bb=0 0 864 504]{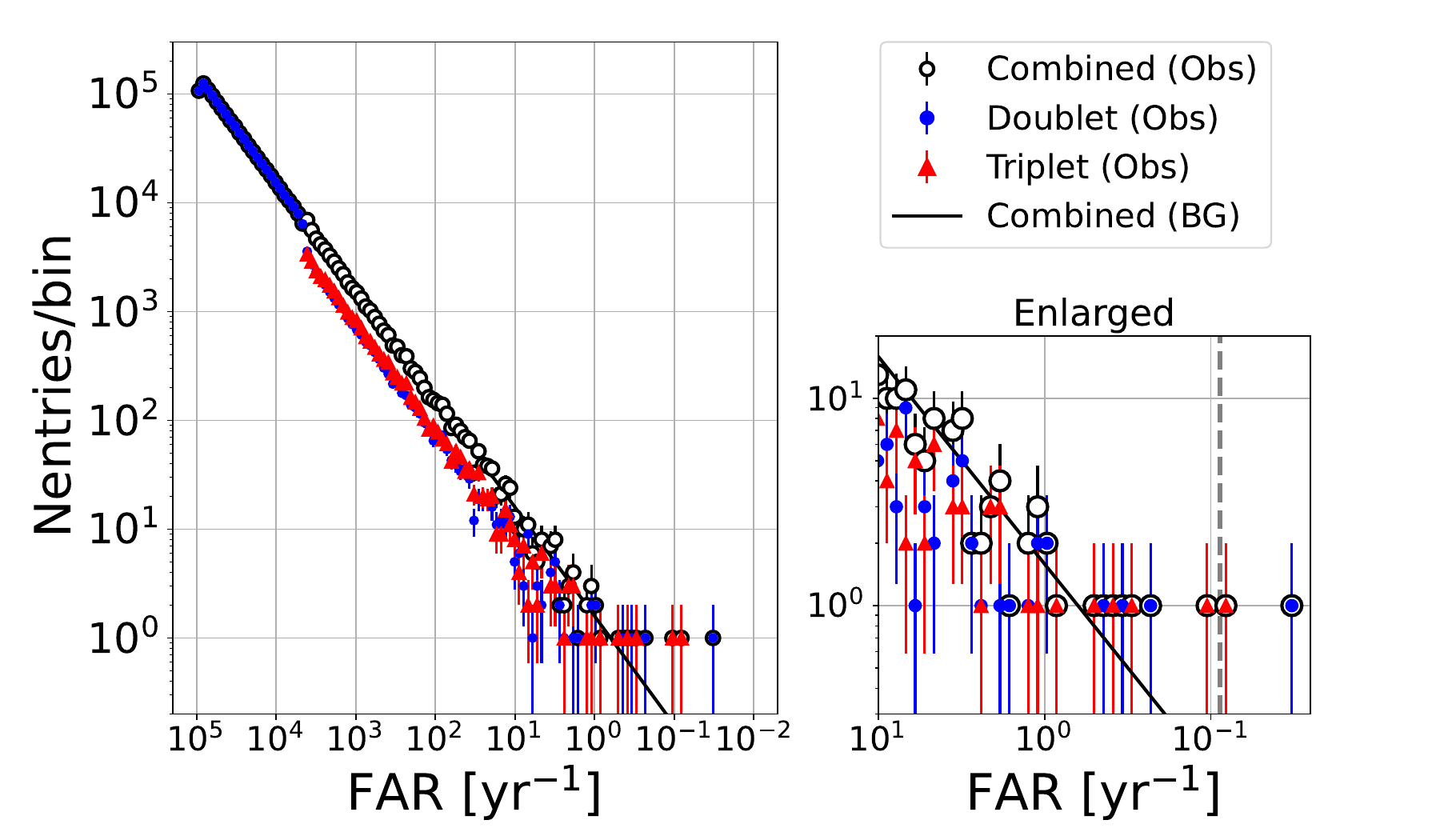}
  \caption{Distribution of the TS-score converted into 
  the corresponding FAR. 
  Blue circles, red triangles, and white empty markers represent doublets, 
  triplets, and combined distributions of 
  experimental data by the TS-score, 
  while the black line represents the expectation from backgrounds. 
  The right panel provides an 
  enlarged view of the large significance region. 
  The vertical dashed line in the right panel corresponds to the inverse of live time of $1/11.4~\mathrm{yr}^{-1}$.}\label{fardist}
\end{figure}

We search for multiplet signals in the IceCube Northern Sky GFU sample
taken from May 2011 to December 2022 after finalizing the selection scheme 
and confirming its performance following a blinded analysis strategy.
No statistically significant doublets or triplets of astrophysical origins are observed.
Figure~\ref{fig_tsdist} shows the $\Lambda$ distributions of doublets and triplets 
showing excellent agreement with the background distributions.
For illustrative purposes, we also display the expected distributions by 
standard-candle neutrino sources with $\mathcal{E}^*_\nu=10^{52}~\mathrm{erg}$ and 
$\rho_{\nu}=10^{-8}~\mathrm{Mpc}^{-3}\,\mathrm{yr}^{-1}$.
Figure~\ref{fardist} shows the distribution of FAR values converted from the TS-score.

\begin{table*}[th]
\begin{center}
\caption{Summary of the top 12 significant multiplets observed in the dataset.\label{eachmult}}
\hspace{-3.6cm}\scalebox{0.9}[0.9]{
\begin{tabular}{l c c c c c c l l l c c c} 
\hline\hline 
Index & Type\mytablenotemark{a} & Fit RA \mytablenotemark{b}& Fit DEC \mytablenotemark{b}
& $\mathrm{log}_{10}E_1$ & $\mathrm{log}_{10}E_2$ & $\mathrm{log}_{10}E_3$ & ~~$\mathrm{MJD}_{1}$\mytablenotemark{c} & ~~$\mathrm{MJD}_{2}$\mytablenotemark{c} & ~~$\mathrm{MJD}_{3}$\mytablenotemark{c} & $\Delta T$\mytablenotemark{d} & Local \mytablenotemark{e}& FAR\mytablenotemark{e}  \\ 
~~~$k$ && (deg) & (deg) & (GeV) & (GeV) & (GeV) & ~~(day) &  ~~(day)& ~~(day) & (day) & p-value  & $(\mathrm{yr}^{-1})$    \\ 
&&&&&&&&&&  &$\times 10^{6}$ &\\
\hline 
~~~1 & D & $110.05$ & $11.05$ & 6.77 & 3.75 & - & 56819.20 & 56813.38 & - & 5.8 & $0.32$ & 0.034   \\
~~~2 & T & $0.58$ & $-0.35$ & 3.62 & 4.31 & 5.47 & 59027.66 & 59015.46${}^+$ & 59011.22* & 16.4 & $0.74 $ & 0.078    \\
~~~$3$ & T & $200.53$ & $6.30$ & 3.59 & 4.34 & 4.26 & 56487.55 & 56459.68${}^\sharp$ & 56458.53${}^\dagger$ & 29 & $1.1 $ & 0.112    \\
~~~4 & D & $0.58$ & $-0.35$ & 4.31 & 5.47 & - & 59015.46${}^+$ & 59011.22* & - & 4.2 & $2.3 $ & 0.30    \\
~~~$5$ & T & $200.54$ & $6.29$ & 3.69 & 4.34 & 4.26 & 56479.74 & 56459.68${}^\sharp$ & 56458.53${}^\dagger$ & 21.2 & $2.9 $ & 0.305   \\
~~~6 & D & $121.15$ & $-2.01$ & 3.94 & 3.97 & - & 59260.68 & 59255.52 & - & 5.2 & $3.1 $ & 0.326    \\
~~~7 & T & $0.55$ & $-0.32$ & 3.67 & 4.31 & 5.47 & 59041.22 & 59015.46${}^+$ & 59011.22* & 30 & $3.8 $ & 0.396    \\
~~~8 & D & $133.75$ & $52.87$ & 4.45 & 4.48 & - & 58759.56 & 58755.30 & - & 4.3 & $4.5 $ & 0.469    \\
~~~9 & T & $219.29$ & $12.79$ & 4.61 & 3.7 & 5.04 & 56817.03 & 56809.62 & 56794.31 & 22.7 & $4.7 $ & 0.487   \\
~~~$10$ & T & $123.38$ & $8.20$ & 3.77 & 4.76 & 4.46 & 59274.42 & 59271.67${}^\ddagger$ & 59257.61${}^\flat$ & 16.8 & $7.6 $ & 0.796  \\
~~~$11$ & D & $200.53$ & $6.29$ & 4.34 & 4.26 & - & 56459.68${}^\sharp$ & 56458.53${}^\dagger$ & - & 1.2 & $8.8 $ & 0.918   \\
~~~$12$ & D & $123.38$ & $8.20$ & 4.76 & 4.46 & - & 59271.67${}^\ddagger$ & 59257.61${}^\flat$ & - & 14.1 & $9.5 $ & 0.987  \\
\hline 
\end{tabular}
}
\end{center}
\vspace{-2mm}
{\bf \small Notes.}
{
\mytablenotetext{a}{\footnotesize The notation of D or T represents doublet and triplet, respectively.}\vspace{-2mm}
\mytablenotetext{b}{\footnotesize The right ascension and declination indicate the most probable direction determined by the fit. }\vspace{-2mm}
\mytablenotetext{c}{\footnotesize The superscripts indicate events that are shared by other multiplets.} \vspace{-2mm}
\mytablenotetext{d}{\footnotesize The time difference $\Delta T$ is the time interval between the first and the last neutrino event. }\vspace{-2mm}
\mytablenotetext{e}{\footnotesize The local p-values are introduced in the main text and converted to the corresponding FAR per year. }\vspace{-2mm}
}
\end{table*}

Table~\ref{eachmult} summarizes the 12 most significant 
multiplets observed in the dataset 
indexed according to their significance by $k$.
We select these multiplets with predetermined thresholds 
of $\mathrm{FAR}=1~\mathrm{yr}^{-1}$. 
\sy{
Because some doublets and triplets share the same events,
the 12 multiplets point to seven unique directions in the Northern Sky.
Note that each local p-value listed in the table is 
obtained directly from the PDF of the TS score. 
The post-trial global p-value is obtained by $1-(1-p_\mathrm{local})^N$, 
that is, the probability that at least one doublets or triplets 
has a local p-value smaller than $p_\mathrm{local}$ in the $N$~multiplet pool.
The most significant multiplet ($k=1$) has a post-trial global p-value
of 0.32, corresponding to a FAR of $0.034\ {\rm yr}^{-1}$.
}

\sy{To probe any statistical deviation from the background-only hypothesis in the dataset as a whole,}
we perform a binomial test (see \hyperref[sec:appendix]{Appendix}) 
for the 12 multiplets, and the p-value evaluated by the pseudo-experiment is 0.38.
Thus, for the entire dataset, the observation is 
consistent with the background-only hypothesis.

\begin{figure}[tbp]
  \hspace{1mm}\includegraphics[width=0.9\linewidth,bb=0 0 720 648]{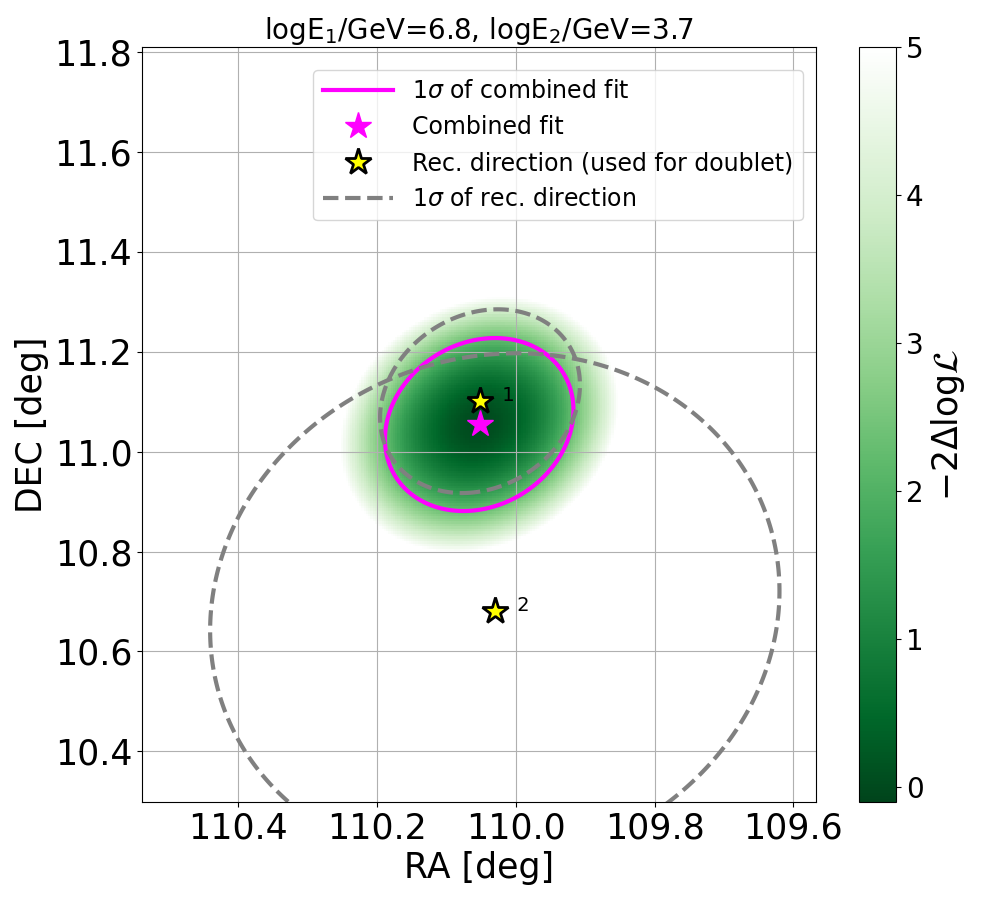}\\
  \includegraphics[width=0.92\linewidth,bb=0 0 720 648]{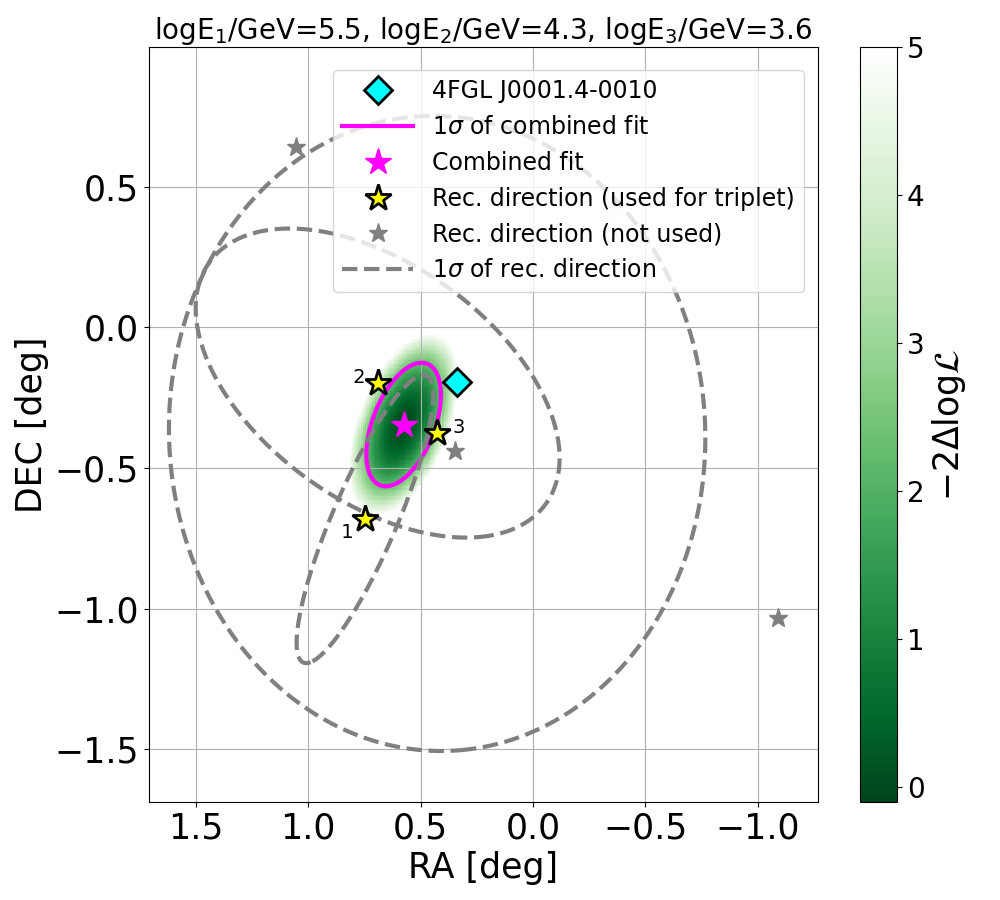}\\
  \caption{The contour plots of the direction of the two most significant multiplets. 
  The upper and lower panels represent
  the doublet ($k=1$) and the triplet ($k=2$), respectively.
  Yellow stars indicate the directions of neutrino events to 
  form the multiplets, while magenta stars denote the fitted directions of the
  multiplets. The dashed lines outline 68\% containment uncertainty regions 
  of each event determined by assuming Gaussian distribution, 
  while magenta solid lines represent that of the fitted directions. 
  The background color scales represent the residuals of the signal likelihoods: $-2 \Delta \log \mathcal{L}^\mathrm{sig}$.
  Small gray stars in the bottom panel represent neutrino 
  events observed around the time of multiplet detection 
  (from MJD59007 to MJD59047), but \sy{are not included in the multiplet found by the seed event method (see Sec.~\ref{clustering_sec})}. 
  In the lower panel, the cyan diamond represents the direction of a 
  Fermi source 4FGL~J0001.4-0010~\citep{ballet2023fermi_DR4}, 
  which is separated by $0.28^{\circ}$ from the best fit direction.}\label{mult_contour}
\end{figure}

Figure~\ref{mult_contour} illustrates the contours 
of the top two significant multiplets $k=1$ and $k=2$, as shown in the upper and lower panels, respectively.
The $k=1$ multiplet emerges as a significant doublet, because one of the 
reconstructed energies is very high at 6~PeV (previously reported in~\cite{Aartsen_2016_kloppo, PhysRevLett.117.241101_EHE2016}).

\begin{figure}[tbp]
  \begin{center}
   \includegraphics[width=\linewidth,bb=0 0 648 576]{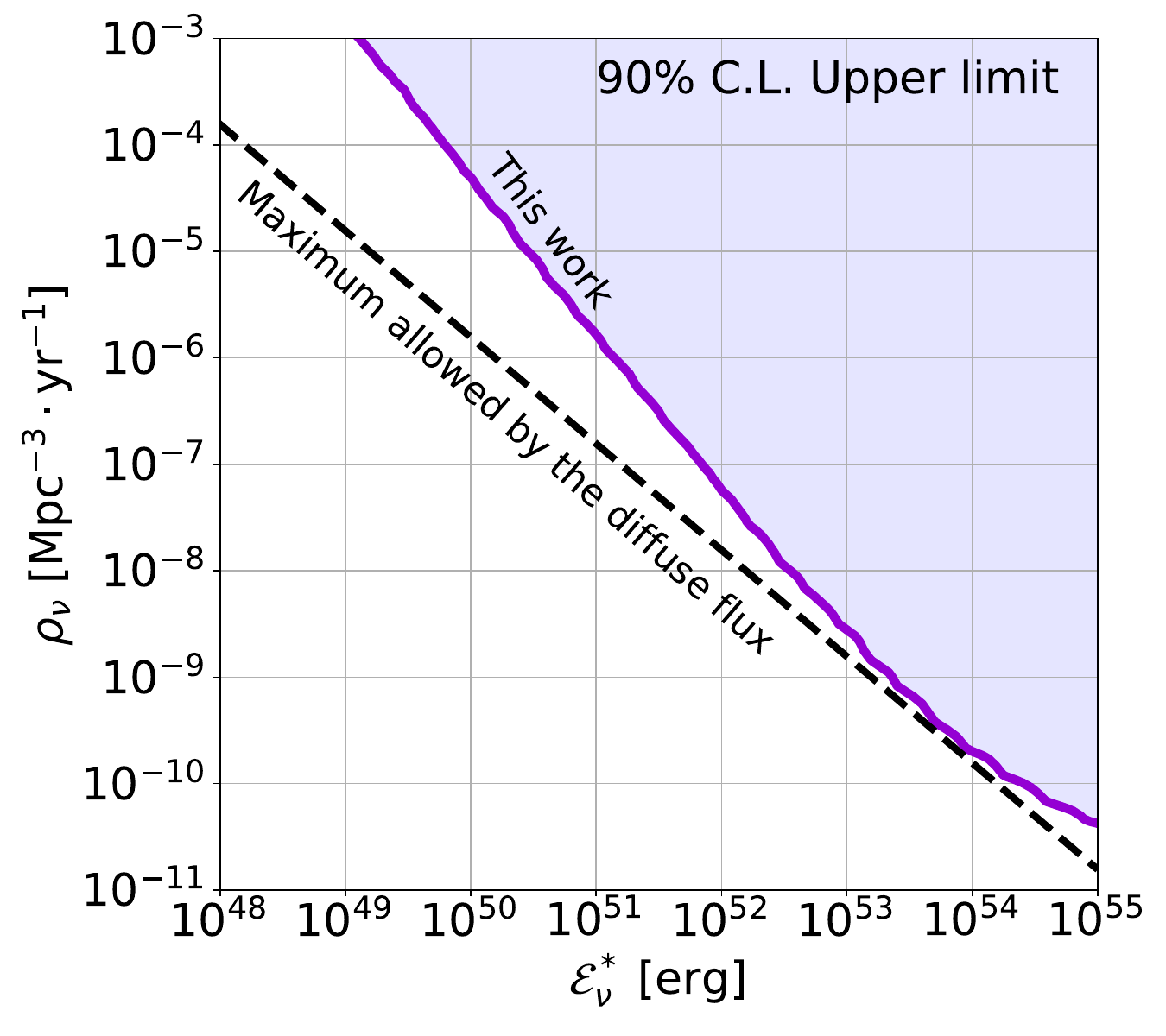}
  \end{center}
 \caption{Contours of excluded region of neutrino transient source 
 parameters ($\mathcal{E}^*_{\nu}$, $\rho_\nu$) 
 at 90\% confidence level when 
 $\Delta T_\mathrm{max}=30~\mathrm{days}$, and 
 the spectral indices are 2.3.
 The area above the curves is excluded at the 90\% confidence level. 
 The diagonal lines correspond to upper limits implied from the diffuse flux when 
 the signal source spectrum is $\gamma=2.3$. 
 These diffuse flux constraints are determined so as not 
 to overshoot the energy distribution of the 
 measured diffuse neutrino flux~\citep{Abbasi_2022} (see Sec.~\ref{subsec:constraintsByDiffuse}). 
 All the constraints are obtained assuming the SFR compatible evolution.
 }
  \label{figCL}
\end{figure}

\def\thefootnote{\fnsymbol{footnote}}

\sy{
The null detection of statistically significant multiplets constrains the parameters of 
$\mathcal{E}^*_{\nu}$ and $\rho_\nu$. We use the largest $\Lambda = 23.5$ ({\it i.e.}, 
the highest inconsistency with the background-only hypothesis) found in the dataset, 
considering that a multiplet exceeding this value favors an astrophysical origin.
For given $(\mathcal{E}^*_{\nu}, \rho_\nu)$, we inject 
neutrinos with a power law of $E^{-2.3}$ from 10~TeV to 1~PeV and zero elsewhere, and calculate 
how often it records a multiplet larger than this value in the observation.
A set of parameters with a frequency higher than 0.9 is ruled out at a 90 \% confidence level.
}
Figure~\ref{figCL} shows the 
contours of the neutrino transient source parameters ($\mathcal{E}^*_{\nu}$, $\rho_\nu$) 
at the 90\% confidence level for the \sy{ $E^{-2.3}$ spectrum.
The area above the curve is excluded at the 90\% confidence level. }

\section{Discussion \label{discussion}}

\begin{figure}[tbp]
  \centering
  \includegraphics[width=\linewidth,bb=0 0 720 504]{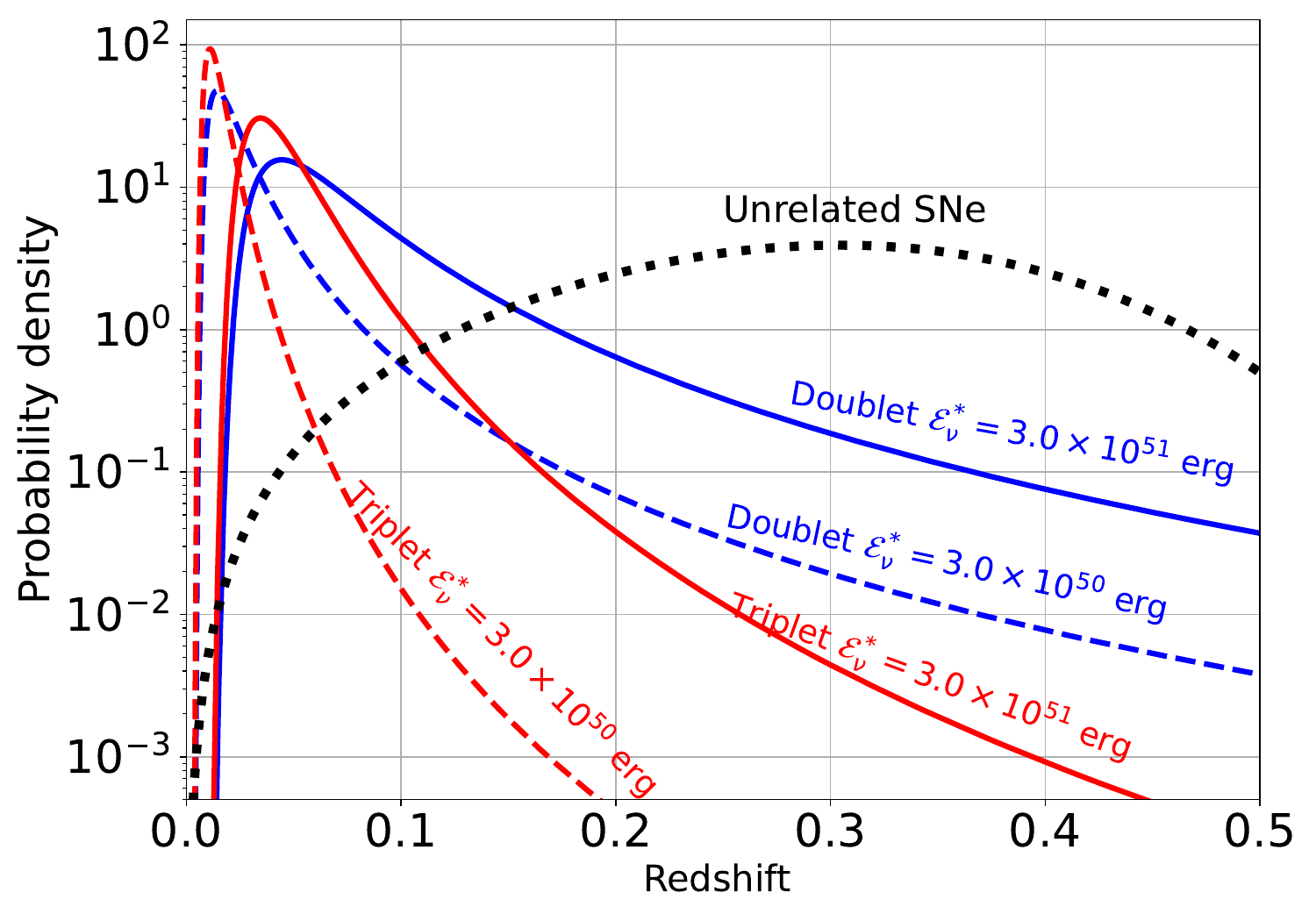}
  \caption{Probability density of the redshift distributions of neutrino sources 
  detected as 30~days doublet (blue) and triplet (red) 
  in the direction of the observed triplet ($k=2$ in Table~\ref{eachmult}) 
  for neutrino emission energies of 
  $3 \times 10^{50}~\mathrm{erg}$ (dashed) and $3 \times 10^{51}~\mathrm{erg}$ (solid). 
  The black dotted curve represents the distribution of the unrelated closest SNe
  with a burst rate density of $10^{-4}~\mathrm{Mpc}^{-3}\,\mathrm{yr}^{-1}$ observed 
  during 30~days and a solid angle of $\Delta \Omega_\nu = \pi\cdot (\mathrm{0.3}^\circ)^{2}$~\citep{Yoshida_2022}.}\label{redshift}
\end{figure}

\subsection{Relationship with constraints by the diffuse neutrino flux\label{subsec:constraintsByDiffuse}}

Transient sources should also contribute to the diffuse neutrino 
flux and their contributions must be less than 
the total observed diffuse flux intensity, resulting 
in parameter constraints on $(\mathcal{E}^*_{\nu}, \rho_\nu)$~\citep{Murase:2016gly, Yoshida_2022}.
The diagonal line in Fig.~\ref{figCL} corresponds to the 
maximally allowed limit of the partial contributions of transient source fluxes 
implied from the diffuse neutrino flux assuming $\gamma=2.3$. 
Note that this value is towards the lower range of spectral indices inferred from the 
diffuse flux by IceCube. 
The upper bound is determined by requiring that the assumed flux does not exceed 
the IceCube track limit (as a benchmark of the flux, we are using the flux level $\phi_{@100~\mathrm{TeV}}^{\nu_{\mu}+\bar{\nu}_{\mu}}=1.44\times 10^{-18}~\mathrm{GeV}^{-1}\,\mathrm{cm}^{-2}\,\mathrm{sec}^{-1}\,\mathrm{sr}^{-1}$ 
and $\gamma=2.37$~\citep{Abbasi_2022}). 
With present sensitivity, the constraint by the multiplet 
analysis is weaker than the interpretation of the measured diffuse flux. 


\subsection{Posterior analysis of higher-rank multiplets}

As mentioned in Sec.~\ref{result_sec}, several multiplets share 
common neutrino events. In general, 
when more than three neutrino events form a multiplet, 
the subsets can become multiple doublets.
In particular, multiplets numbered $k=2,7$ and $k=3,5$ 
formed quadruplets within 30 days. 
The significance of the quadruplets are 
calculated a posteriori. 
The TS is similarly extended to 
higher-rank multiplets and the local p-value is defined as an upper 
percentile of the TS distribution in the scrambled dataset. 
The two quadruplets have post-trial global p-values of 0.57 and 0.27, respectively, 
and they are 
consistent with backgrounds. Moreover, 
these are the two most significant quadruplets in the entire dataset.


\subsection{Multi-messenger astronomy using the neutrino multiplets \label{sec_MM}}

Searches for the astrophysical counterparts of neutrino multiplets 
by multi-messenger observations are a powerful strategy for
the identification of high-energy neutrino transient sources. 
It has been pointed out~\citep{Eddinton_biassingdist,Yoshida_2022} that
neutrino multiplet detections effectively limit the detectable distance of 
neutrino sources, thereby provides a chance to evaluate a physical association.
For example, a posteriori, we find that a Fermi source 
4FGL~J0001.4-0010~\citep{Abdollahi_2022_fermicatalog, ballet2023fermi_DR4} is 
within 90\% of containment uncertainty 
region of the $k=2$ triplet 
(within $0.3^\circ $ of the fitted direction, see Fig.~\ref{mult_contour}). 
The host galaxy FBQS~J0001-0011 was identified as a BL~Lac ~\citep{AAquasarcatalog} 
with its redshift of $z=0.46$~\citep{2017ApJS..233...25A}.
The probability of random coincidence in finding a Fermi source within this region 
is estimated to be approximately 5\% based on the 
local number density of the Fermi sources. 
Given that the typical density of BL~Lac is $O(10^{-7})~\mathrm{Mpc}^{-3}$~\citep{Ajello_2014}, 
the effective rate density in 30~days is $\rho_{\nu} \sim 10^{-6}~\mathrm{Mpc}^{-3}\, \mathrm{yr}^{-1}$ which gives constraint on 
$\mathcal{E}^*_\nu < 3\times 10^{51}~\mathrm{erg}$ from Fig.~\ref{figCL}.
Figure~\ref{redshift} shows the PDF of the redshift 
of transient sources for two cases $\mathcal{E}^*_\nu = 3\times 10^{50}$~erg and $3\times 10^{51}$~erg.
The observed redshift of FBQS~J0001-0011 
is unlikely as a source of the triplet.
Similarly, even if other frequent transients 
such as type-Ia SNe (as shown by the dotted curve) 
are observed in the direction of the observed multiplets 
by other follow-up studies, they are also unlikely if $z \gtrsim 0.1$.

It has been argued that bright TeV-PeV neutrino sources could be related to 
bright X-ray sources with a high opacity to $\gamma$-rays~\citep{PhysRevLett.116.071101}.
Motivated by this scenario, we perform a correlation analysis using publicly available data 
recorded by the Monitor of All-sky X-ray Images (MAXI)~\citep{MAXI_10.1093/pasj/61.5.999}. 
The X-ray signals were detected using Gas Slit Cameras (GSC) with a sensitive energy range 
from 2~keV to 20keV. The point spread function of the GSC is approximately 1 degree. 
\sy{The operation period entirely covers the range when the data in the present analysis is taken.}
We focus on X-ray photons ranging from 4~keV to 10~keV due to the 
high detection efficiency of GSCs and 
a relatively low background rate~\citep{MAXI_energy_range_10.1093/pasj/63.sp3.S677}.
The noise detected in GSCs originates from accidental hits of 
charged particles and the rate depends on 
the time. To cancel out time-dependent noise, 
we define the signal and control regions in the vicinities of the neutrino multiplet directions 
as a $1.5^{\circ}$~circle and a $3.0^{\circ}$~doughnut shape with 
inner radius of $1.5^{\circ}$, respectively, and 
calculate the X-ray flux as the difference between these regions. 
The statistical uncertainty of the flux is estimated 
by propagating Poisson fluctuations of the photon counts in both regions. 
No significant X-ray emission is observed beyond the noise level
in association with the detected neutrino multiplets, and thereby 90\% CL upper 
limits on the average X-ray energy flux over $\Delta T$
are placed in each of the multiplets except for $k=8$ and $k=12$ (no X-ray data available).
The upper limits of the X-ray energy flux is approximated as \vspace{0mm}
\begin{align}
F_{\rm X}&<8\times 10^{-11}\left(\frac{\Delta T}{\mathrm{day}}\right)^{-1/2}~\mathrm{erg}\,\mathrm{cm}^{-2}\,\mathrm{s}^{-1}
\label{eq:FxUL}
\end{align}
at the 90\% CL.

\section{Conclusion \label{conclusion_sec}}

\sy{
We present a search for high-energy neutrino multiplets on the timescale
of 30~days using 11.4 years of IceCube data.
Dedicated search algorithms and likelihood constructions
are developed to optimize the month-scale transient phenomena.
No significant doublets or triplets are observed, providing
constraints on bright but rare source populations. 
The upper limit is weaker than the constraint based on the 
measured diffuse flux by the IceCube.
}

Future multi-messenger observations using neutrino multiplet detections 
as well as high statistical searches brought about by future neutrino detectors 
such as IceCube-Gen2~\citep{IceCube-Gen2:2020qha}
are promising for understanding high-energy neutrino emission mechanisms.


\section*{Acknowledgements}

The IceCube collaboration acknowledges the significant 
contributions to this manuscript from Nobuhiro Shimizu.
The authors gratefully acknowledge the support from the following agencies and institutions:
USA {\textendash} U.S. National Science Foundation-Office of Polar Programs,
U.S. National Science Foundation-Physics Division,
U.S. National Science Foundation-EPSCoR,
U.S. National Science Foundation-Office of Advanced Cyberinfrastructure,
Wisconsin Alumni Research Foundation,
Center for High Throughput Computing (CHTC) at the University of Wisconsin{\textendash}Madison,
Open Science Grid (OSG),
Partnership to Advance Throughput Computing (PATh),
Advanced Cyberinfrastructure Coordination Ecosystem: Services {\&} Support (ACCESS),
Frontera computing project at the Texas Advanced Computing Center,
U.S. Department of Energy-National Energy Research Scientific Computing Center,
Particle astrophysics research computing center at the University of Maryland,
Institute for Cyber-Enabled Research at Michigan State University,
Astroparticle physics computational facility at Marquette University,
NVIDIA Corporation,
and Google Cloud Platform;
Belgium {\textendash} Funds for Scientific Research (FRS-FNRS and FWO),
FWO Odysseus and Big Science programmes,
and Belgian Federal Science Policy Office (Belspo);
Germany {\textendash} Bundesministerium f{\"u}r Bildung und Forschung (BMBF),
Deutsche Forschungsgemeinschaft (DFG),
Helmholtz Alliance for Astroparticle Physics (HAP),
Initiative and Networking Fund of the Helmholtz Association,
Deutsches Elektronen Synchrotron (DESY),
and High Performance Computing cluster of the RWTH Aachen;
Sweden {\textendash} Swedish Research Council,
Swedish Polar Research Secretariat,
Swedish National Infrastructure for Computing (SNIC),
and Knut and Alice Wallenberg Foundation;
European Union {\textendash} EGI Advanced Computing for research;
Australia {\textendash} Australian Research Council;
Canada {\textendash} Natural Sciences and Engineering Research Council of Canada,
Calcul Qu{\'e}bec, Compute Ontario, Canada Foundation for Innovation, WestGrid, and Digital Research Alliance of Canada;
Denmark {\textendash} Villum Fonden, Carlsberg Foundation, and European Commission;
New Zealand {\textendash} Marsden Fund;
Japan {\textendash} Japan Society for Promotion of Science (JSPS)
and Institute for Global Prominent Research (IGPR) of Chiba University;
Korea {\textendash} National Research Foundation of Korea (NRF);
Switzerland {\textendash} Swiss National Science Foundation (SNSF).

\section*{Appendix Binomial p-value }\label{sec:appendix} 
Suppose that there are $N$ local p-values $\{p_1, p_2,\ldots, p_N\}$ that are 
sorted in ascending order (from the smallest to the largest): $\{q_1, q_2,\ldots q_N\}$. 
Under the background-only hypothesis, $k$-th smallest p-value, $x=q_k$, is distributed as \vspace{-1mm}
\begin{align}
\rho_k(x) = k \begin{pmatrix}
 N \\ k 
\end{pmatrix} x^{k-1}(1-x)^{N-k}, \label{binomialeq1}
\end{align}
such that the p-value of each $q_k$ is given as a cumulative distribution function, as in \vspace{1mm}
\begin{align}
P(k)= \int_0^{q_k} \hspace{-1mm}\rho_k(x)\,dx = k \begin{pmatrix}
 N \\ k 
\end{pmatrix} B_{q_k}(k, N-k+1),
\end{align}
where $B_x(a, b)=\int_0^x t^{a-1}(1-t)^{b-1} dt$ is an incomplete beta function.
By integrating by parts, this can be expressed in a different form: 
\begin{align}
P(k)=\sum_{m=k}^{N} \begin{pmatrix}
 N \\ m 
\end{pmatrix} q_k^m (1-q_k)^{N-m}, \label{bonomialeq2}
\end{align}
and are sometimes referred to as binomial p-values. Note that these quantities are {\it trial-corrected} ({\it i.e.,} $P(k)$ 
is uniformly distributed in $(0,1)$ if $p_k~(k=1,2,\ldots,N)$ is also uniformly distributed).
In particular, from Eq.~(\ref{bonomialeq2}), 
the p-value of the most significant observation ($q_1$) is calculated as $P(1)=1-(1-q_1)^N$, which is the trial-corrected 
global p-value introduced in Sec.~\ref{result_sec}. 

To define the significance of the entire dataset, 
the smallest binomial p-value $\min{P(k)}$ in subsets $\{P(1), P(2),\ldots, P(M)\}$ ($M=12$ as shown in Table~\ref{eachmult}) can be used. 
This $\min{P(k)}$ cannot be interpreted as a p-value, thus 
the distribution of $\min{P(k)}$ is numerically evaluated by pseudo-experiments, 
and a lower percentile of the distribution is computed to define the p-value for the observed $\min{P(k)}$.

\bibliography{main}{}
\bibliographystyle{aasjournal}



\end{document}